\documentclass{article}
\usepackage[margin=1in]{geometry}

\usepackage{amsmath,amssymb, bm}

\usepackage{changepage}

\usepackage[utf8x]{inputenc}

\usepackage{cite}

\usepackage{nameref,hyperref}

\usepackage[table]{xcolor}

\usepackage{array}

\usepackage{alltt}

\usepackage{float}
\usepackage[lined,boxed]{algorithm2e}

\usepackage{color}
\definecolor{light-gray}{gray}{0.95}
\usepackage{listings}
\usepackage[prependcaption,textsize=small]{todonotes}
\usepackage[]{color}
\usepackage[]{graphicx}

\usepackage{framed}
\makeatletter
\newenvironment{kframe}{%
 \def\at@end@of@kframe{}%
 \ifinner\ifhmode%
  \def\at@end@of@kframe{\end{minipage}}%
  \begin{minipage}{\columnwidth}%
 \fi\fi%
 \def\FrameCommand##1{\hskip\@totalleftmargin \hskip-\fboxsep
 \colorbox{white}{##1}
 \hskip-\fboxsep
     \hskip-\linewidth \hskip-\@totalleftmargin \hskip\columnwidth}%
 \MakeFramed {\advance\hsize-\width
   \@totalleftmargin\z@ \linewidth\hsize
   \@setminipage}}%
 {\par\unskip\endMakeFramed%
 \at@end@of@kframe}
\makeatother

\usepackage{authblk}

\title{\texttt{sourceR}: Classification and Source Attribution of Infectious Agents among Heterogeneous Populations} 

\author[1]{Poppy Miller} 
\author[2,3]{Jonathan Marshall}
\author[3,4,5]{Nigel French}
\author[4]{Chris Jewell}

\affil[1]{CHICAS, Faculty of Health and Medicine, Lancaster University, Lancaster,  UK}
\affil[2]{Institute of Fundamental Sciences, Massey University, Palmerston North, New Zealand}
\affil[3]{mEpiLab, Massey University, Palmerston North, New Zealand}
\affil[4]{New Zealand Food Safety Science and Research Centre}
\affil[5]{New Zealand Institute for Advanced Studies}

\begin{document}
\maketitle

\begin{abstract} 
Zoonotic diseases are a major cause of morbidity, 
and productivity losses in
both humans and animal populations. Identifying the source of food-borne zoonoses (e.g. an animal reservoir or
food product) is crucial for the identification and prioritisation of food safety interventions. For many zoonotic
diseases it is difficult to attribute human cases to sources of infection because there is
little epidemiological information on the cases. However, microbial strain typing allows zoonotic
pathogens to be categorised, and the relative frequencies of the strain types among the
sources and in human cases allows inference
on the likely source of each infection. We introduce \texttt{sourceR}, an \texttt{R} package for
quantitative source attribution, aimed at food-borne diseases. 
 It implements a
fully joint Bayesian model using strain-typed surveillance data from both human
cases and source samples, capable of identifying important sources of infection. The model measures the force of infection from each
source, allowing for varying survivability, pathogenicity and virulence of pathogen
strains, and varying abilities of the sources to act as vehicles of infection. A Bayesian
non-parametric (Dirichlet process) approach is used to cluster pathogen strain types by
epidemiological behaviour, avoiding model overfitting 
and allowing detection of strain types associated with potentially high 'virulence'. 

\texttt{sourceR} is demonstrated using
\emph{Campylobacter jejuni} isolate data collected in New Zealand between 2005 and 2008.
Chicken from a particular poultry supplier was identified
as the major source of campylobacteriosis which is qualitatively similar 
to results of previous studies using the same dataset.
Additionally, the software identifies a cluster of 9 MLSTs with abnormally high
'virulence' in humans.

\texttt{sourceR} enables straightforward attribution of cases of zoonotic infection to putative sources of infection by epidemiologists and public health decision makers. As 
\texttt{sourceR} develops, we intend it to become an important and flexible resource for food-borne disease attribution studies.
 
 \end{abstract}
 
\section{Introduction}
\subsection{Background}
Zoonotic diseases are a major source of human morbidity world wide. In 2010, there were an estimated 600 million cases globally \cite{HavEtAl2015}, of which 96 million were \emph{Campylobacter spp.} (resulting in 21 thousand deaths \cite{WHOdb15}). Attributing cases of food-borne disease to putative sources of infection is crucial to identify and prioritise food safety interventions. Traditional approaches to source attribution include full risk assessments, analysis and extrapolation of surveillance or outbreak data, and
analytical epidemiological studies \cite{CrGrAn02}. However, their results can be highly uncertain due to long and variable disease incubation times, and many and various 
exposures of an individual to potential sources of infection. Given this difficulty, quantitative methods using pathogen strain type frequency have shown promise for statistically identifying important sources of food-borne 
illness \cite{MulJonNob09}.  

For a given disease, quantitative source attribution uses typing of pathogen isolates from human cases and suspected sources of infection (food and environmental).   
Samples are screened for the presence of the pathogen, with isolates then categorised using a typing methodology. Multilocus Sequence Typing (MLST) is a commonly used genotyping method providing a relatively coarse characterisation of isolates of bacterial species \cite{UrMai03}. An MLST sequence type is defined as a unique combination of alleles at several gene loci, typically located in conserved regions of the genome \cite{DingColWar01}.

Routine surveillance for food-borne pathogens is now commonplace in many countries and is performed by national authorities, for example FoodNet in the US \cite{Allos15042004}, the Danish 
Zoonosis Centre (\url{food.dtu.dk}), and the Ministry for Primary Industries in New Zealand (\url{foodsafety.govt.nz}).  Despite this availability of data we are unaware of any previous implementations in standard statistical software for source attribution modelling, with past analyses being performed using a variety of \emph{ad hoc} methodologies.  Moreover, current statistical source attribution models have strong assumptions, computational approximations or inherent identifiability problems (discussed further in the `Review of models and notation' section).

This paper presents an \texttt{R} package \texttt{sourceR}, which implements a flexible Bayesian non-parametric model, designed for use by epidemiologists and public health decision makers to attribute cases of 
zoonotic infection to putative sources of infection.  We first describe a motivating example and review previous source attribution models before describing our model innovations, demonstrating the software, and discussing results and future directions.

\subsection{Motivating example}\label{sec:motivation}

In 2006, New Zealand had one of the highest incidences of campylobacteriosis in 
the developed world, with an annual incidence in excess of 400 cases per 100,000 people \cite{BakWilIkr06}.  The data set was first published in \cite{MuelColMid10}, with a detailed description of the data (and data collection methods) available in \cite{FreMar09} and \cite{FreMar13}.  A campaign to change poultry processing procedures, supported in part by results from previous
quantitative source attribution approaches, was successful in leading to a sharp decline in campylobacteriosis incidence after 2007 \cite{MulJonNob09}.

The data consists of MLST-genotyped \emph{Campylobacter} isolates (from both human cases of campylobacteriosis and potential food and environmental sources) collected
between 2005 and 2008 in the Manawatu region of New Zealand. These data are included in our \texttt{sourceR} package (named \texttt{campy}). We use this data set as a case study, and compare our results with previously published statistical approaches.

\subsection{Review of models and notation}\label{sec:review}

In this section we define our notation, and briefly review the approaches that have been used previously to analyse \texttt{campy}.  For a given time period, we denote by $y_i$ the number of human cases of a disease caused by pathogen type $i=1,\dots,n$.  For the same time period, we let $s_j$ denote the total number of source samples collected from source $j=1,\dots,m$, for which $x_{ij}$ are positive for pathogen type $i$.

\subsubsection{Hald model}\label{sec:hald}

The approach of Hald \emph{et al.} \cite{HaldVosWed04} was to compare the number of human cases caused by different pathogen types with their prevalence in different food sources (whilst accounting for type and source specific effects). This requires a heterogenous distribution of pathogen types among the food sources. The number of human cases for each type $y_i$ is modelled as a Poisson random variable with mean given by a linear combination of source specific effects, type specific effects and source sample \emph{Campylobacter} contamination prevalences.

\begin{eqnarray}
y_i & \sim & \mbox{Poisson} (\lambda_i) \\
\lambda_i & = & q_i\sum_{j=1}^{m} \alpha_j c_j p_{ij}\label{eq:haldmodel}
\end{eqnarray}
where for source $j$ $c_j$ is the annual exposure, $p_{ij}=r_{ij}\times k_j$ is the absolute prevalence of each pathogen type with $k_j = \sum_{i=1}^{n} x_{ij}/s_j$ the prevalence of positive samples and $r_{ij}=\frac{x_{ij}}{\sum_{i=1}^{n}x_{ij}}$ the relative prevalence of each pathogen type.

The unknown parameters in the model are the vectors $\bm{q}$ and $\bm{\alpha}$.  Here, $\bm{q}$ represents the characteristics that determine a type's capacity to cause an infection (such as survivability during food processing, pathogenicity and virulence), and $\bm{\alpha}$ accounts for the ability of a particular source to act as a vehicle of infection.  These parameters are interpreted further in \nameref{S4_type_source_effects}. Inference is performed in a Bayesian framework allowing the model to explicitly include and quantify the uncertainty surrounding each of the parameters. 

Equation \ref{eq:haldmodel} over-specifies the model, with $m+n$ parameters (the source and type effects) but only $n$ independent observations (the observed human case totals $y_{i}$). In the original approach \cite{HaldVosWed04}, identifiability 
was obtained by \emph{a priori} clustering of the elements of $\bm{\alpha}$ and $\bm{q}$. In constrast, the Modified Hald model \cite{MulJonNob09} prefers to reduce the effective number of parameters by treating $\bm{q}$ as a log Normal$(0, \tau)$ distributed random effect. However, a strong prior is needed on $\tau$ to shrink $\bm{q}$ towards 0 sufficiently to avoid overfitting the model, the choice of which is arbitrary.

The Modified Hald model introduces uncertainty into the relative prevalence matrix by modelling the source sampling process. This model was fitted in WinBUGS using an approximate two stage process \cite{MulJonNob09}. First, a posterior distribution was estimated for the absolute prevalence of source types $\bm{p}
$, using the model specified in Eqs \ref{eq:mhrij} and \ref{eq:mhpij}
:
\begin{eqnarray}
r_{\cdot j}\sim\textsf{Dirichlet}(\bm{1}) \; \forall \; j \label{eq:mhrij} \\
k_{j}\sim\textsf{Beta}(1,1) \; \forall \; j \label{eq:mhpij}
\end{eqnarray}

   The marginal posterior for each element of $\bm{p}$ was then approximated by a Beta distribution
$$
p_{ij}\sim\textsf{Beta}(w_{ij},v_{ij})
$$
using the method of moments to calculate $w_{ij}$ and $v_{ij}$. These were used as independent priors for each $p_{ij}$ 
which removes the constraint that they sum to $k_{j}$ over each type $i$.  Thus, the absolute prevalence for source $j$ ($\sum_{i=1}^{I}p_{ij}$) is no longer constrained to be a probability (as it may be larger than 1). 

\subsubsection{Asymmetric Island model}

The Asymmetric Island Model \cite{WilGabLea08, iSource} takes a different approach to the models described above.  Here, the evolutionary processes (mutation, migration and recombination) of the 
sequence types are modelled to infer probabilistically the source of each human infection using genetic data from each subtype. The extra information in the genetic typing allows the model to attribute human cases from a type not observed in any sources to a likely source of infection by comparing the genetic similarity to other types that are observed in the sources. This is not possible with the Hald or Modified Hald models, however, they are much simpler with fewer assumptions and a wider range of suitable data (for example, phenotypic typing can be used).  We include results from this model as a comparison in the `Results' section.

\section{Design and Implementation} \label{sec:methods}

Our approach addresses the problems inherent in both the Hald and Modified Hald models. We introduce a fully joint model for both source and human case 
sampling allowing us to integrate over uncertainty in the source sampling process, estimating both the prevalence of contaminated source samples and the relative prevalence of each identified 
type (without resorting to an approximate marginal probability distribution on $\bm{p}$).
Furthermore, we introduce non-parametric clustering of pathogen types using a Dirichlet Process (DP) model on 
the type effect vector $\bm{q}$, providing an automatic data-driven way of reducing the effective number of parameters to aid model identifiability.  We are able, therefore, to circumvent the Hald model 
requirement for heuristically grouping pathogen types, as well as avoiding an arbitrary prior distribution specification for the random effect precision parameter ($\tau$) required by 
the Modified Hald model.

Often, human case data is associated with location such as urban/rural, or even GPS coordinates.  On the other hand, food samples are likely to be less spatially constrained due to distances between production and sale locations.  
Also, both human and source data may exist for multiple time-periods.  We therefore denote the number of human cases of time $i$ occurring in time-period $t$ at location $l$ by $y_{itl}$, the number of samples of source $j$ in time-period $t$ by $s_{jt}$, with the type counts $x_{ijt}$.  We allow for different exposures of humans to sources in different locations, by allowing the source effects to vary between times and locations, $\alpha_{jtl}$.

\subsection{Model} \label{sec:model}

As with the Hald model, we assume the number of human cases $y_{itl}$ identified by isolation of subtype $i$ in time-period $t$ at location $l$ is Poisson distributed
\begin{equation}\label{eq:likelihood}
y_{itl}\sim \textsf{Poisson}(\lambda_{itl} = q_{i} \sum_{j=1}^{m} \alpha_{jtl} p_{ijt})
\end{equation}

For each source $j$, we model the number of positive source samples
\begin{equation}
\bm{x}_{jt} \sim \mbox{Multinomial}(s^{+}_{jt}, \bm{r}_{jt})
\end{equation}
where $\bm{x}_{jt} = \left(x_{ijt}, i=1,...,n\right)^T$ denotes the vector of type-counts in source $j$ in time-period $t$, $s^{+}_{jt} = \sum_{i=1}^{n} x_{ijt}$ denotes the number of positive samples obtained, and $\bm{r}_{jt}$ denotes a vector of relative prevalences $Pr\left(\textsf{type}_{i} | \textsf{source}_{j}, \textsf{time}_{t} \right)$.  The advantage of this model is that it automatically places the constraint $\sum_{i=1}^{n} r_{ijt} = 1$,  avoiding the approximation made in \cite{MulJonNob09} where 
independent Beta-distributed priors were assigned marginally to components of $\bm{r}_{jt}$.  The source case model is then coupled to the human case model through the simple relationship 
\begin{equation}
p_{ijt} = r_{ijt}k_{jt}
\end{equation}
where $k_{jt}$ is the prevalence of any isolate in source $j$ in time-period $t$.

In principle, a Beta distribution could be used to model $k_{jt}$, arising as the conjugate posterior distribution of a Binomial sampling model for $s^{+}_{jt}$ positive samples from $s_{jt}$ tested, 
and a Beta prior on $k_{jt}$. We instead choose to fix the source prevalences at their empirical estimates 
($k_{jt}=s^{+}_{jt}/s_{jt}$) because the number of source samples is typically high. 

The type effects $\bm{q}$, which are assumed invariant across time or location, are drawn from a DP with base distribution $Q_0$ and a concentration parameter $a_q$
\begin{equation}
q_i \sim \mbox{DP}\left(a_q, Q_0 \right). \label{eq:qDP}
\end{equation}
The DP groups the elements of $\bm{q}$ into a finite set of clusters $1:\kappa$ (unknown \emph{a priori}) with values $\theta_1,...,\theta_\kappa$ meaning bacterial types are clustered into groups with similar epidemiological behaviour.

Heterogeneity in the source matrix $\bm{x}$ is absolutely required to identify clusters from sources, which may not be guaranteed \emph{a priori} due to the observational nature of the data collection.

\subsection{Inference}
This section describes how the model is fitted in a Bayesian context by first describing the McMC algorithm used to fit this model, then developing the prior model.

\subsubsection{McMC algorithm} \label{mcmc_section}
The joint model over all unobserved and observed quantities is fitted using Markov chain Monte Carlo (McMC, full details in \nameref{S6_mcmc_alg}). The source effects and relative prevalence parameters are updated using independent adaptive Metropolis-Hastings updates \cite{RobRos06}. 
The type effects $\bm{q}$ are modelled using a DP (Eq \ref{eq:qDP}) with a Gamma base distribution
$Q_0 \sim \mbox{Gamma}(a_\theta, b_\theta)$.  As the Gamma distribution is conjugate
with respect to the Poisson likelihood (Eq \ref{eq:likelihood}), it is possible to use a marginal Gibbs sampler within a Polya Urn, or ``Chinese restaurant process'' construction \cite{GelCarSte13} (see \nameref{S6_mcmc_alg}). This was chosen over the more general ``Stick breaking process'' because it allows sampling from the conditional posterior of $\bm{\theta}$.  This is particularly important when the elements of $\bm{\theta}$ values are highly dispersed: a base distribution with little mass near the locations of some of the true values for the groups results in poor mixing for the group allocations using the stick breaking algorithm (as it is difficult for a type to change group when no other groups have a suitable $\theta$ value). In contrast, the marginal scheme allows an element of $\bm{q}$ to move into a new cluster, then samples a $\theta$ value directly from the conditional posterior for that group, improving group mixing dramatically.  

\subsubsection{Priors} \label{priors_section}

The source and type parameters ($\bm{\alpha}_{tl}$ for all $t$ and $l$, and $\bm{q}$ respectively) account for a multitude of source and type specific factors which are difficult to quantify \emph{a priori}.  Therefore, with no single real-world interpretation, the distributional form of the priors were chosen for their flexibility. A Dirichlet prior is placed on each $\mathbf{r}_{jt}$ which suitably constrains its L1 norm, i.e. $\sum_{i=1}^n r_{ijt} = 1$. A Dirichlet prior is also placed on each
$\bm{\alpha}_{tl}$, with the constrained L1 norm aiding identifiability between the mean of the source and type effect parameters. For more detail on specifying parameters for the Dirichlet Process and priors see the \nameref{S2_DP}.

\subsection{Code implementation}

Standard McMC packages (e.g. WinBUGS, Stan, PyMC3) all lack the capability to implement marginal Gibbs sampling for Dirichlet processes, necessitating a custom McMC framework (see section `Extensibility').  We chose R as a platform because of its ubiquity in epidemiology, and advanced support for post-processing of McMC samples.  Minimal dependencies on other R packages are required, and are installed automatically. 

\texttt{sourceR} uses an object-oriented design, which allows separation of the model from the McMC algorithm.  Internally, the model is represented as a directed acyclic graph (DAG, see \nameref{S3_DAG}) in which nodes are represented by an R6 class hierarchy.  Generic adaptive Metropolis Hastings algorithms are attached to each parameter node, with the conditional independence properties of the DAG allowing automatic computation of the required (log) conditional posterior densities. 

A difficulty with the DAG setup is the representation of the Dirichlet process model on the type effects $\bm{q}$, since each update of the marginal Gibbs sampler requires structural alterations.  Therefore, we subsume the entire Dirichlet process into a single node, with a bespoke marginal Gibbs sampling algorithm written for our Gamma base-distribution and Poisson likelihood model.

\section{Results}  \label{sec:results}
The case study below (using the \texttt{campy} (campylobacteriosis) data set described in the `Motivation' section) illustrates how the \texttt{sourceR} package is used in practice to identify important sources of infection. We compare the results of our Bayesian non-parametric approach with results from the Modified Hald and Asymmetric Island models, and additionally the historical `Dutch' model (see \nameref{S1_dutch} and \cite{Dutch1999}). The priors for our model were selected to be non-informative. 

The prevalence $k_j$ is calculated by dividing the number of positive samples 
by the total number of samples for each source.  In the data below, we note that for several samples the MLST typing failed, with the number of positive samples exceeding the apparent total number of MLST-typed isolates. However, assuming MLST typing fails independently of pathogen type, this does not bias our results. 

The work flow for fitting the model begins with removing types with no source cases and calculating the prevalences $k$. 
\begin{lstlisting}
data(campy)
zero_rows <- which(apply(campy[, c(2 : 7)], 1, sum) == 0)
campy <- campy[-zero_rows,]
total_samples = c(239, 196, 127, 595, 552, 524)
positive_samples = c(181, 113, 109, 97, 165, 86)
k <- data.frame(Value = positive_samples / total_samples,
                Source = colnames(campy[, 2:7]),
                Time = rep("1", 6),
                Location = rep("A", 6))
\end{lstlisting}
The data and model parameters are set using the \texttt{HaldDP\$new()} constructor. Starting values are selected automatically unless provided via a list named \texttt{init} to the constructor.
\begin{lstlisting}
priors <- list(a_theta = 0.01, b_theta = 0.00001, 
               a_alpha = 1, a_r = 0.1)
my_model <- HaldDP$new(data = campy, k = k, priors = priors, a_q = 0.1)
\end{lstlisting}
McMC control parameters are be passed via \texttt{fit\_params}
\begin{lstlisting}
my_model$fit_params(n_iter = 1000, burn_in = 10000, thin = 500)
\end{lstlisting}
The model is run using the \texttt{update} function. Additional iterations may be appended using \texttt{append = TRUE}.
\begin{lstlisting}
set.seed(59623)
my_model$update()
# my_model$update(n_iter = 10000, append = T)
\end{lstlisting}
We provide the \texttt{extract} method for ease of access to the complex posterior.
\begin{lstlisting}
## returns the posterior for the r, alpha, q, c, 
## lambda_i, lambda_j and lambda_j_prop parameters
my_model$extract()
\end{lstlisting}
The \texttt{extract} function returns the posterior for the selected parameters as a list with a multidimensional array for each of \texttt{alpha}, \texttt{r}, \texttt{q}, \texttt{s}, \texttt{lambda\_j} and \texttt{lambda\_i}.

\begin{figure}
\centering{
\includegraphics[width=0.8\textwidth]{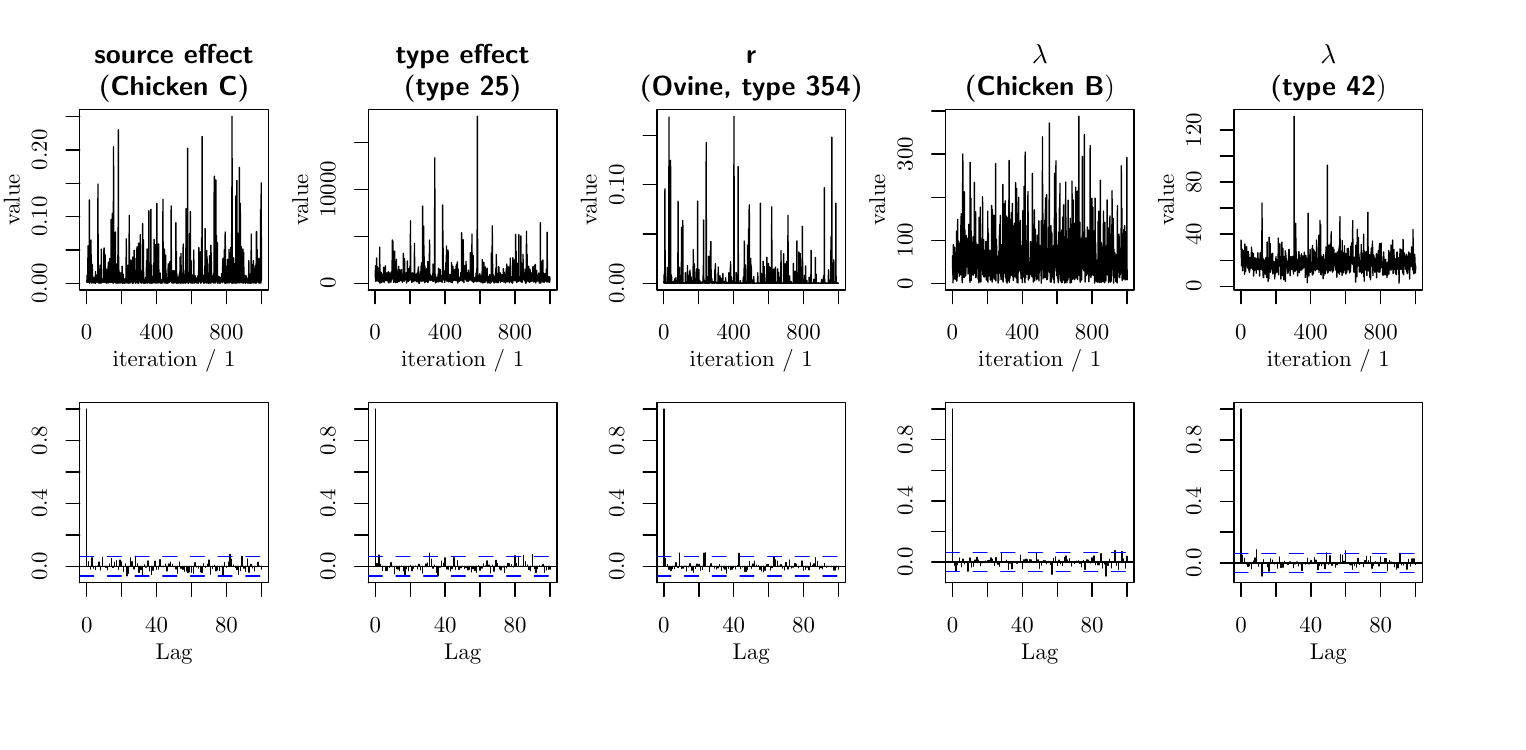} 
\caption{Trace and acf plots for a sample of the model 
parameters (Manawatu \emph{Campylobacter} data).}\label{fig:trace_params_real}}
\end{figure}

\begin{figure}
\centering{
\includegraphics[width=0.8\textwidth]{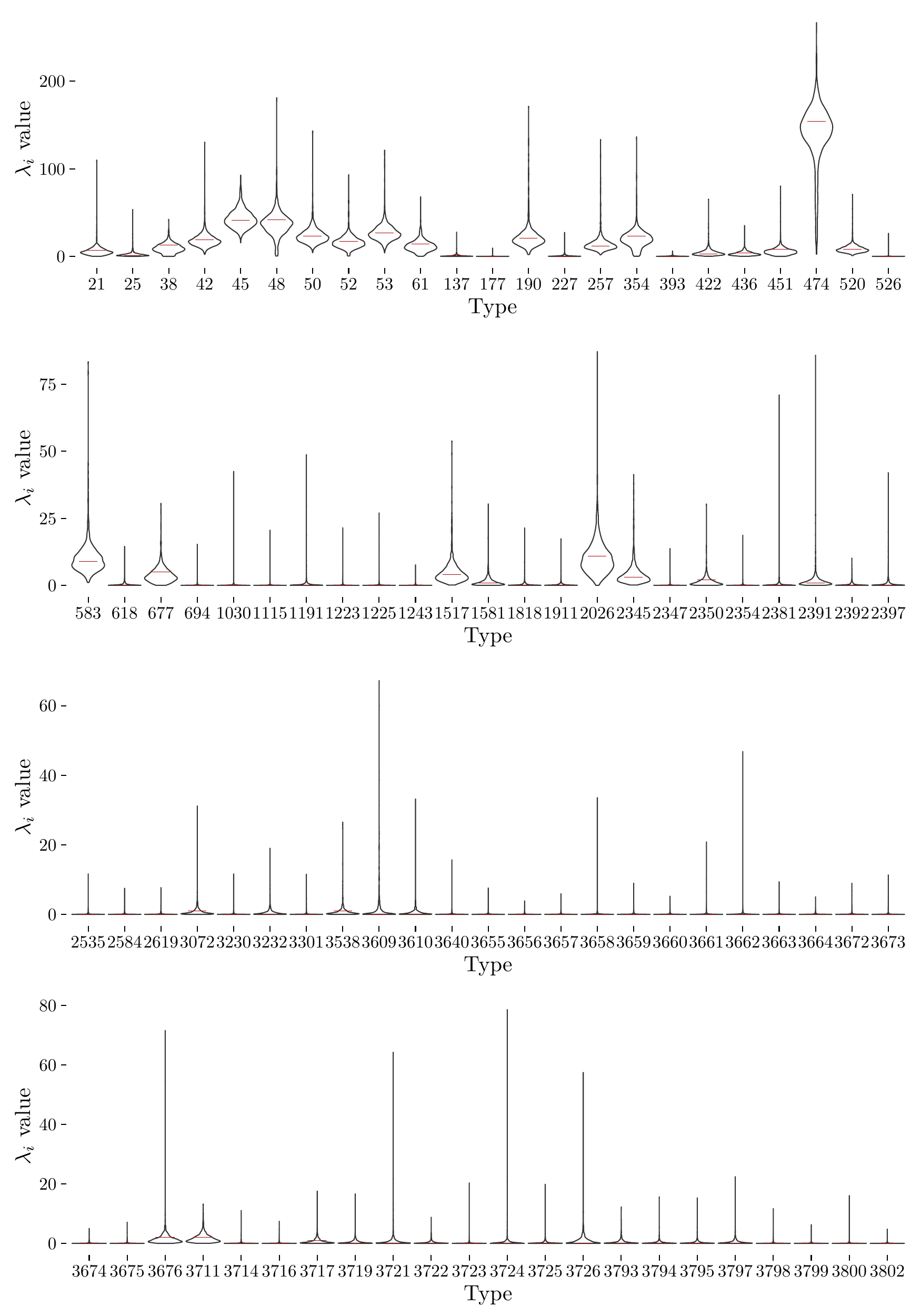} 
\caption{Violin plots showing the marginal posteriors for each $\lambda_i$ (number of cases attributed to each type). Observed number of cases for each type are shown as horizontal red lines. (Manawatu \emph{Campylobacter} data).}\label{fig:lambda_i_residuals_real}}
\end{figure}

Trace and autocorrelation plots for the parameters indicate that the Markov chain is mixing well and has converged, and that thinning by 500 is adequate (Figure~
\ref{fig:trace_params_real}). The residual plots for the $\lambda_{i}$s (Figure~\ref{fig:lambda_i_residuals_real}) show that the model fits well.
\begin{lstlisting}
## Plot the marginal posteriors for the following parameters
## source effect for Chicken supplier C
plot(my_model$extract(params="alpha", sources="ChickenC")$alpha, type="l")
## type effect 25
plot(my_model$extract(params="q", types="25")$q, type="l")
## relative prevalence for source effect Ovine, type 354
plot(my_model$extract(params="r", sources="Ovine", types="354")$r, 
     type="l")
## number of cases attributed to Chicken supplier B
plot(my_model$extract(params="lambda_j", sources="ChickenB")$lambda_j, 
     type="l")
## number of cases attributed to sub type 42
plot(my_model$extract(params="lambda_i", types="42")$lambda_i, type="l")
\end{lstlisting}
The \texttt{summary()} function calculates medians and credible intervals calculated with three possible methods (percentile, SPIn \cite{LiuGelZhe15}, or Chen-Shao \cite{ChenShao99}).
\begin{lstlisting}
my_model$summary(alpha = 0.05, CI_type = "percentiles")
\end{lstlisting}
These can be used to plot an observation versus fitted plot as follows
\begin{lstlisting}
## The summary for lambda i is a 4D array with
## [type, time, location, (median/ CI_lower/ CI_upper)]
med_li_vals <- my_model$summary(alpha = 0.05, params = "lambda_i", 
                time = "1", location = "A")$lambda_i[, , , "median"]
human_cases <- my_model$print_data()$y
plot(med_li_vals, human_cases)
\end{lstlisting}

\begin{figure}
\centering{
\includegraphics[angle =0, width=\textwidth]{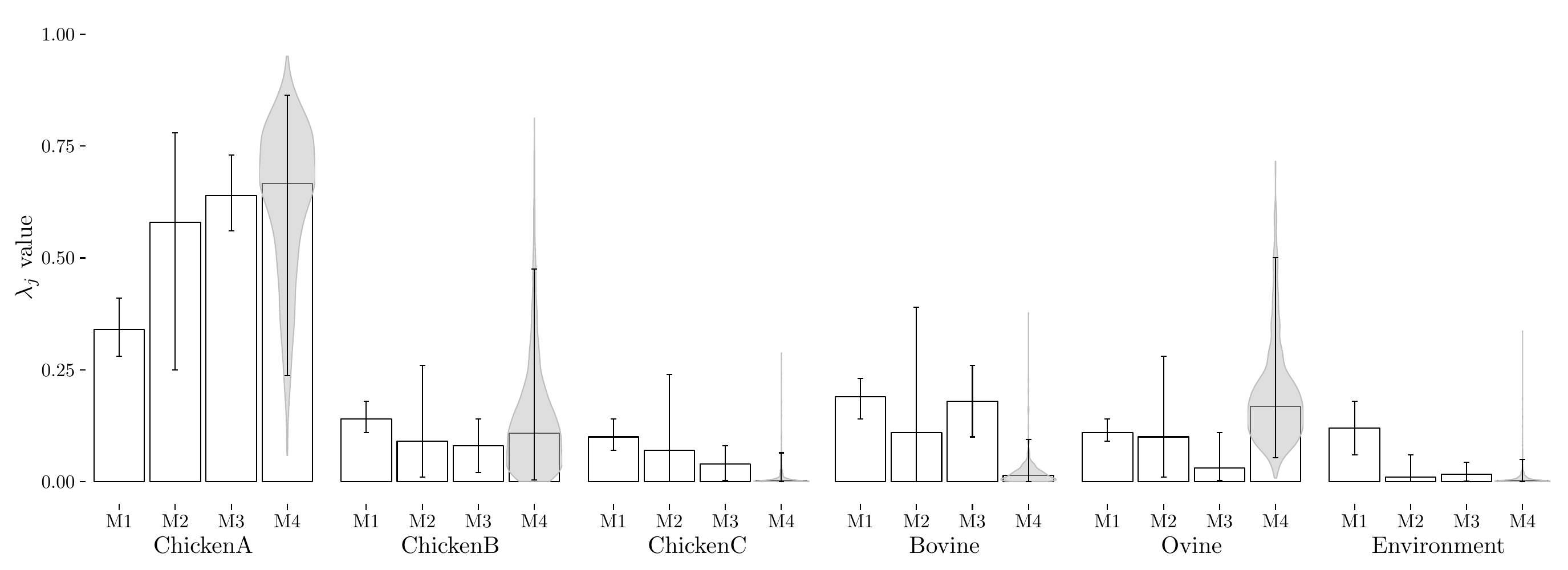} 
\caption{Comparison of the proportion of human campylobacteriosis cases 
attributable to each source for the models: M1 (Dutch model), M2 (Modified Hald model), M3 (Island model) and M4 (HaldDP model). Error bars represent 95\% confidence or credible intervals. Violin plots show the marginal posteriors of the $\lambda_j$ parameters for the HaldDP model.}\label{fig:lambda_j_real}}
\end{figure}

\begin{kframe}
\end{kframe}\begin{figure}
\centering{
\includegraphics[width=0.8\textwidth]{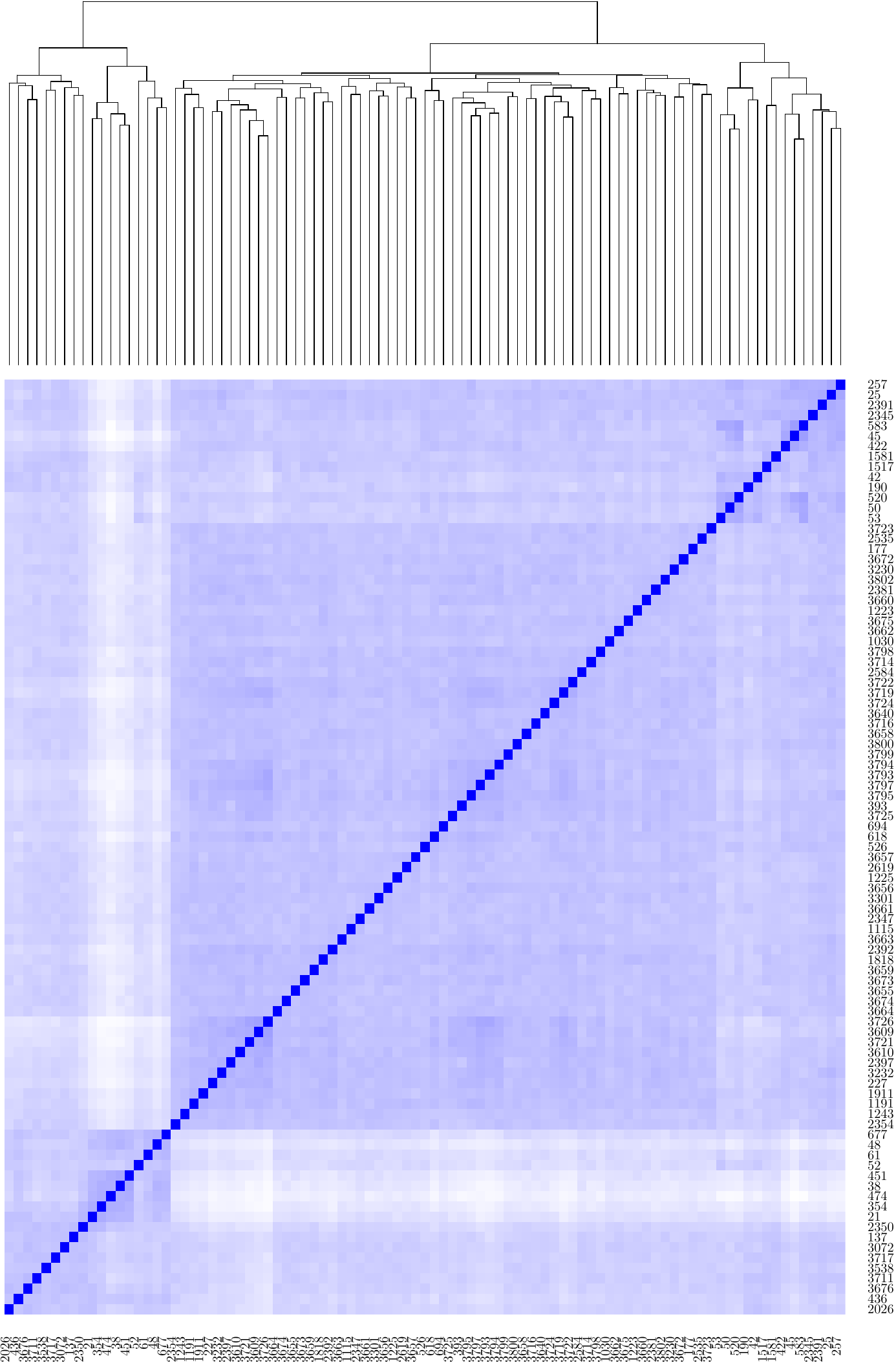} 
\caption{Heatmap showing the grouping of the type effects (q) using the Manawatu \emph{Campylobacter} data. A white pixel represents a dissimilarity value of 1 between a pair of sub types, whilst dark blue (see pixels on the diagonal) gives a value of zero.}
\label{fig:type_effect_heatmap_real}}
\end{figure}

\begin{figure}
\centering{
\includegraphics[width=0.8\textwidth]{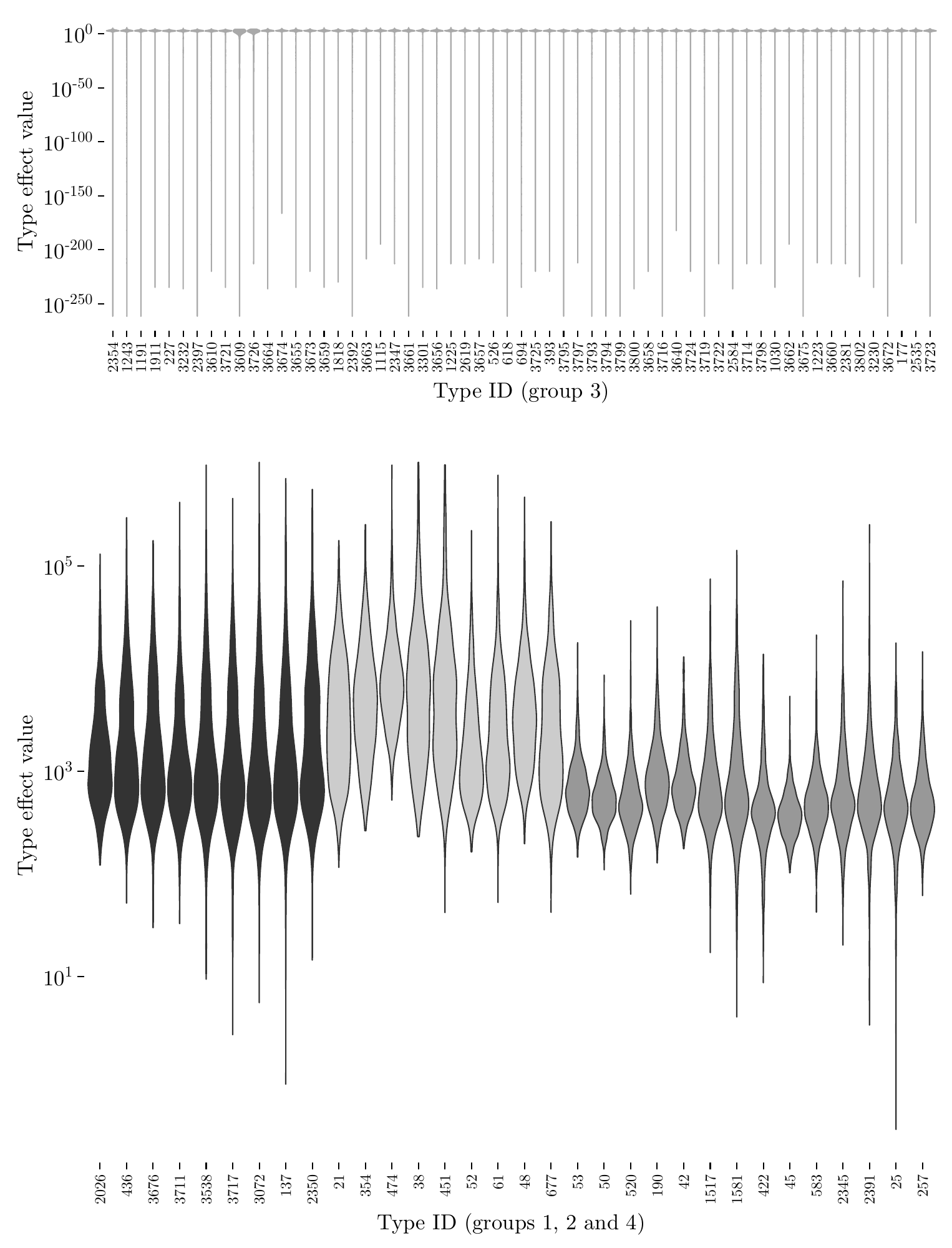} 
\caption{Violin plots of the marginal distributions of the type effects (q) using the Manawatu \emph{Campylobacter} data (using a log scale axis).}\label{fig:type_effect_violinplots_real}}
\end{figure}

See \nameref{S5_running_package} and \nameref{S8_sim_study} for more details on using the package.

\subsection{Type effect marginal distributions and clustering}

To visualise how the DP has clustered the type effects, we use Gower's distance \cite{Gow71} to compute a dissimilarity matrix between all possible pairs of types.  To aid interpretation of the posterior clustering of the type effects under the DP, we provide a method \texttt{plot\_heatmap()} that plots the clustering as a heatmap with a dendrogram. Figure~\ref{fig:type_effect_heatmap_real} shows that the DP identified four main pathogen type clusters. 
\begin{lstlisting}
my_model$plot_heatmap()
\end{lstlisting}	
The violin plots of the marginal posterior distributions for each type effect (Figure~\ref{fig:type_effect_violinplots_real}) show the largest group of types has very small type effects. These correspond to types observed in few source samples and no human cases. Consequently, there is very little information for their type effects which results in very wide credible intervals. The other three groups have much larger type effects. The clustering results identify clusters of strains having particular traits that could be explored using further genotyping or phenotyping assays.

\subsection{Comparison of the proportion of cases attributed to each source} \label{sec:compPropnAttrib}
\begin{figure}[H]
\centering{
\includegraphics[width=0.8\textwidth]{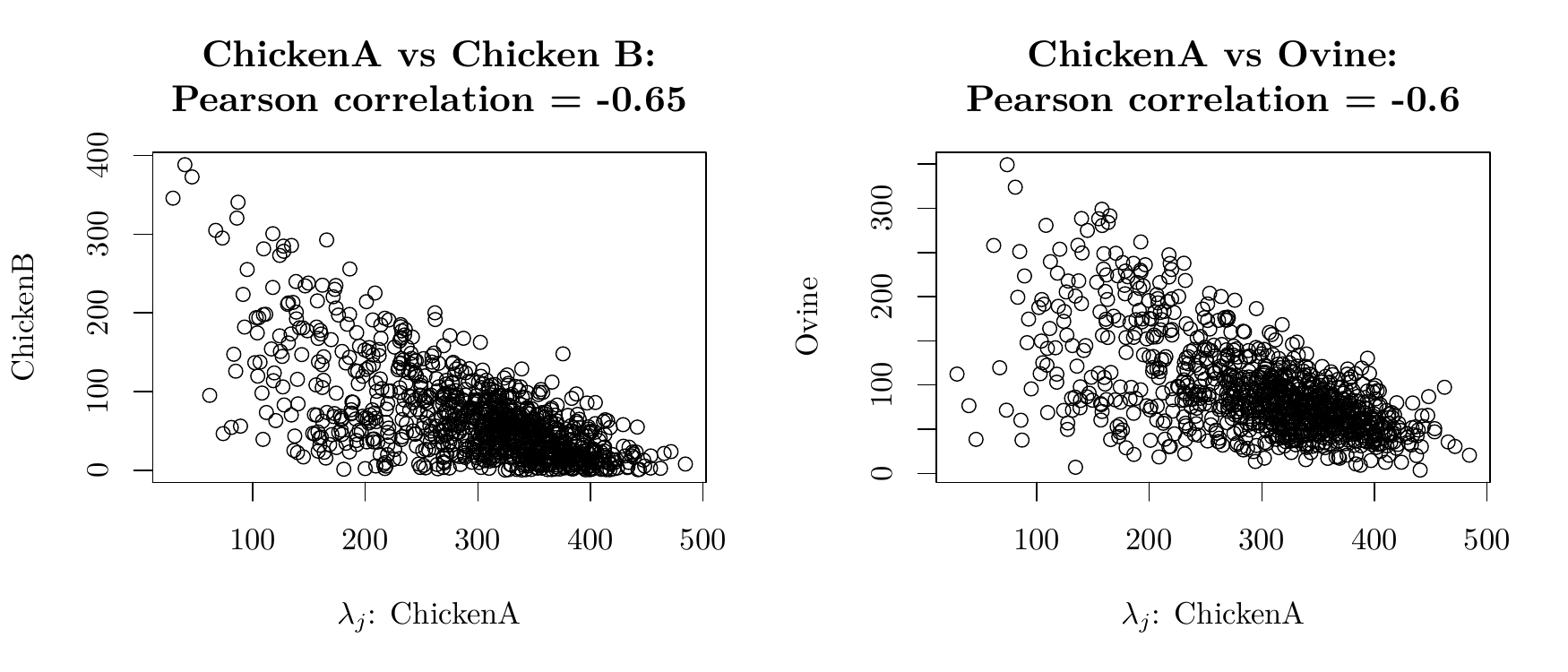} 
\caption{Scatter plots showing the correlation between the $\lambda_j$ marginal posteriors for Chicken supplier A versus Chicken supplier B and Ovine.}\label{fig:cor_plots}}
\end{figure}

Figure~\ref{fig:lambda_j_real} shows the proportion of cases attributed to each source for the HaldDP model and three commonly used source attribution models. The median values are similar between all models except the Dutch method \cite{Dutch1999}. The Dutch model confidence intervals are very narrow because there are far fewer parameters in the model; however, the lack of source and type effects in the model biases the results. The credible intervals produced by the Island model may be narrow due to more accuracy (as additional genetic information is used). The wide credible intervals for the the HaldDP and modified Hald models may be due to C. jejuni’s complex epidemiology resulting in relatively large uncertainty for the disease origin \cite{MulJonNob09}, and posterior correlations between some parameters. In particular, the new model shows that the proportion of cases attributed to poultry supplier A is negatively correlated with the proportion of cases attributed to both ovine and poultry supplier B sources (Pearson correlation coefficients of -0.60 and -0.65 respectively, see Fig \ref{fig:cor_plots} in \nameref{S7_cor_plot}). The HaldDP model gives a more accurate representation of the uncertainty inherent in source attribution. Some of this non-identifiability is not fully explored in the Modified Hald model \cite{MulJonNob09} as fitting the model in two stages does not allow full propagation of the uncertainty. In particular, when calculating the hyper-parameters for the Beta priors for each $p_{ij}$ from the first stage model, the authors imposed a minimum $\alpha_{ij}$ of 1. This prevents 'bath tub' shaped beta priors for any $p_{ij}$s which makes the model easier to fit at the expense of discouraging full exploration of the marginal posteriors for $p_{ij}$s that truly have a bath-tub shape. 

\section{Availability and Future Directions}
The stable release version of \texttt{sourceR} is available from the Comprehensive R Archive Network, released under a GPL-3 licence.  The development version is available at \url{http://fhm-chicas-code.lancs.ac.uk/millerp/sourceR}.  As this package develops, we intend \texttt{sourceR} to become a platform for new source attribution model development, providing a central analytic resource for public health professionals.  The establishment of a standard package with a familiar interface will therefore lead to improved repeatability and reusability of source attribution analyses, supporting national public health and hygiene policy decisions.

\subsection{Package extensibility}\label{sec:extensibility}

With increased interest in source attribution models for both food-borne pathogens, and \texttt{sourceR} has been written with extensibility in mind, with the DAG representation allowing for rapid construction of modified or new models.  The package routines are written in R (as opposed to C or C++) to aid readability, with the node class hierarchy and three stage workflow designed to aid the addition of new model classes.  All internal classes and methods are documented to enable prospective developers to familiarise themselves with the source code quickly.  We note that the DAG framework is not limited solely to source attribution models and may used for other Bayesian applications, particularly those for which a Dirichlet process is required.

\subsection{Model extensions}

The main focus of extending \texttt{sourceR} will be on modelling spatiotemporal correlation in the time- and location- dependent parameters.  With the trend in collecting precise geolocation data with human cases, and improved traceability of food, a spatiotemporal correlation model on $\bm{\alpha}_{tl}$ could be used to identify particular foci of source contamination, therefore enabling targeted investigation of particular food supply regions.  Implementation of time varying type effects may be appropriate, particularly in the face of evidence that \emph{Campylobacter} can evolve quickly, with genetic variation conferring virulence not apparent from course-scale MLST typing \cite{WilGabLeath09}.  
Interaction terms between some sources and types would allow for the biologically plausible possibility that certain types are more or less likely to survive and cause disease, dependent on the food source they appear in. This would occur if a specific type was particularly well adapted to a certain food source. However, including interaction terms would significantly increase the number of parameters and reduce identifiability of the model. 

\subsection{Increasing McMC efficiency}
Testing has revealed that the current Metropolis-Hastings based fitting algorithm suffers a loss of efficiency if the source matrix is sparse or highly unbalanced, imbuing negative correlations between certain type/source effect combinations.  Gradient-based fitting algorithms such as Hamiltonian Monte Carlo (HMC) \cite{DuanKenPen87} are designed to converge to high-dimensional, non-orthogonal target distributions much more quickly, and are a target of future development.  In particular,  the No U-Turn Sample (NUTS) presents an attractive method for tuning HMC adaptively, a quality which we consider necessary to minimise user intervention and maximise research productivity \cite{HomGel14}.

\section{Conclusions} \label{conclusion_section}

We have presented a novel source attribution model which builds upon, and unites, the Hald and Modified Hald approaches. It is widely applicable, fully joint, and does not require approximations or a large number of assumptions. Mixing and \emph{a posteriori} correlations are significantly decreased in comparison to the Modified Hald model. Furthermore, it allows the data to inform type effect clustering 
using a Bayesian non-parametric model which identifies groups of bacterial sub types with similar 
putative virulence, pathogenicity and survivability. This is a significant improvement over the previous attempts to improve model identifiability (fixing some source and type effects, or modelling the type effects as random using a 2 stage model). 
Like the Modified Hald model, the new model incorporates uncertainty in the prevalence matrix into the model, however, it does this by fitting a fully joint model rather than a 2 step model. This 
has the advantage of allowing the human cases to influence the uncertainty in the source cases and preserves the restriction on the sum of the prevalences for each source. The \texttt{sourceR} package implements this
model to enable straightforward attribution of cases of zoonotic infection to putative sources of infection by epidemiologists and public health decision makers.

\section{Supporting Information}

\paragraph*{S1 Appendix.}
\label{S1_dutch}
{\bf Dutch model overview} 

The Dutch method \cite{Dutch1999} is one of the simplest models for source attribution. It compares the number of reported human cases caused by a particular bacterial subtype with the relative 
occurrence of that subtype in each source. The number of reported cases per subtype and reservoir is estimated by:
\begin{eqnarray}
\lambda_{ij}=\frac{r_{ij}}{\sum_j r_{ij}}y_i
\end{eqnarray}
where $r_{ij}$ is the relative occurrence of bacterial subtype $i$ in source $j$,
$y_i$ is the estimated number of human cases of type $i$ per year,
$\lambda_{ij}$ is the expected number of cases per year of type $i$ from source $j$. A summation across types gives the total number of cases attributed to source $j$,
denoted by $\lambda_j$:
\begin{eqnarray}
\lambda_j=\sum_i \lambda_{ij} 
\end{eqnarray}
As the Dutch model has no inherent statistical noise model, confidence intervals for the estimated total attributed cases $\hat{\lambda}_j$ by bootstrap sampling over the data set. This model implicitly assumes that there are no source or type specific effects (such as differing virulence of types, or differing consumption of food sources) which is not plausible for most zoonoses. 

\paragraph*{S1 Table.}
\label{S1_Table}
{\bf Summary of model parameters.} 

The following table gives a list of the model parameters for easy reference.
\begin{table}[H] 
\caption{Description and definition of the model parameters.}\label{table_params}
\setlength{\tabcolsep}{4pt}
\bigskip{}
\centering
\label{table:HaldModelParams}
\begin{tabular}{lll}
\multicolumn{1}{l}{\textbf{Parameter}} & \multicolumn{1}{l}{\textbf{Description}} & \multicolumn{1}{l}{\textbf{Estimation}}\\
$\lambda_{ijtl}$ & Number of human cases from type $i$, source $j$ & $\lambda_{ijtl}=\alpha_{jtl}\cdot q_i\cdot r_{ijt}\cdot k_{jt}$\\
 & time $t$ and location $l$&\\
 
$\lambda_{itl}$ & Number of human cases from type $i$ & $\lambda_{itl}=\sum_{j=1}^{m}\lambda_{ijtl}$\\
& time $t$ and location $l$&\\

$\lambda_{jtl}$ & Number of human cases from source $j$ & $\lambda_{jtl}=\sum_{i=1}^{n}\lambda_{ijtl}$\\
& time $t$ and location $l$&\\

$y_{itl}$ & Number of human cases from type $i$ & $y_{i}\sim \textsf{Poisson}(\lambda_{itl})$\\
& time $t$ and location $l$&\\
 &&\\
$x_{ijt}$ & Number of positive samples (that were & Data\\
& successfully MLST typed) from source $j$, type $i$ & \\
&time $t$&\\
$h_{ijt}$  & Number of positive samples (PCR) that & Data\\
& could not be MLST typed. & \\
$s^{+}_{jt}$ & Total number of samples from source $j$ time $t$ & Data\\
 &&\\
$k_{jt}$ & Prevalence of contamination for each source & $\sum_{i=1}^{I}(x_{ijt}+h_{ijt})/s^{+}_{jt}$\\
$r_{ijt}$ & Relative occurrence of type $i$ on source $j$ & $\mathbf{r}_{jt}\sim \textsf{Dirichlet}(a_r)$\\
&time $t$& or $x_{ijt}/\sum_{i=1}^{n}x_{ijt}$\\
$p_{ijt}$ & Absolute prevalence of type $i$ in source $j$ & $r_{ijt}\cdot k_{jt}$\\
&time $t$&\\
$\alpha_{jtl}$ & Unknown source effect for source $j$ & $\mathbf{\alpha}_{tl}\sim \textsf{Dirichlet}(a_{\alpha})$\\
& time $t$ and location $l$&\\
$q_{i}$ & Unknown type effect for type $i$ in group $k$,& $\mathbf{q}\sim \textsf{DP}(\textsf{Gamma}(a_{\theta}, b_{\theta}), a_q)$\\
 & where group $k$ has an unknown value $\theta_k$ &\\
\end{tabular}\\
\end{table}

\paragraph*{S2 Appendix.}
\label{S2_DP}
{\bf Dirichlet Priors and Process details} 

The Dirichlet Process is a random probability measure defined by a base distribution $Q_0$ and a concentration parameter $a_q$ \cite{Fer73}. The base distribution constitutes a prior 
distribution in the values of each element of the type effects $\mathbf{q}$ whilst the concentration parameter encodes prior information on the number of groups $K$ to which the pathogen types are 
assigned. For small values of $a_q$, samples from the DP are likely to have a small number of atomic measures with large weights. For large values, most samples are likely to be distinct, and 
hence, concentrated on $Q_{0}$. A value of 1 implies that, \emph{a priori}, two randomly selected types have probability 0.5 of belonging to the same cluster \cite{GelCarSte13}.

\paragraph{Specifying the Dirichlet Process base distribution and concentration parameters:} The concentration parameter of the DP is specified by the analyst as a modelling decision. The concentration parameter specifies how strong the prior grouping is. In the limit $a\rightarrow 0$, all types will be assigned to one group, increasing $a$ makes a larger number of groups increasingly likely. The Gamma base distribution $Q_{0}$ induces a prior for the cluster locations. This prior should not be too diffuse because if these locations are too spread out, the penalty in the marginal likelihood for allocating individuals to different clusters will be large, hence the tendency will be to overly favour allocation to a single cluster. However, the prior parameters may have a stronger effect than anticipated due to the small size of the relative prevalence and source effect parameters. This can been seen by considering the marginal posterior for $\theta_k$
\[
\theta_{k} \sim \textsf{Gamma}\left(a_{\theta}+\sum_{i:S_{i}=k}y_{i}, b_{\theta}+
\sum_{i:S_{i}=k}\sum_{j=1}^{m}\alpha_{j}\cdot p_{ij}\right)
\]
The term $\sum_{i:S_{i}=k}\sum_{j=1}^{m}\alpha_{j}\cdot p_{ij}$ is very small (due to the Dirichlet priors on $\mathbf{\alpha}$ and $\mathbf{r}_j$), which can result in even a fairly small rate parameter ($b_{\theta}$) dominating.

\paragraph{Specifying Dirichlet priors:}
The simplest Dirichlet priors for the source effects and relative prevalences are symmetric (meaning all of the elements making up the parameter vector $\bm a$ have the same value $a$, called the concentration parameter). Symmetric Dirichlet distributions are used as priors when there is no prior knowledge favouring one component over another. When $a$ is equal to one, the symmetric Dirichlet distribution is uniform over all points in its support. Values of the concentration parameter above one prefer variates that are dense, evenly distributed distributions, whilst values of the concentration parameter below 1 prefer sparse distributions. Note, a prior of 1 for the relative prevalences is too strong (if a relatively non-informative prior is preferred) when there are many observed zero's in the source data; a prior value of 0.1 is more suitable. A more informative prior can be specified by using a non-symmetric Dirichlet distribution. The magnitude of the vector of $\bm a$ parameters corresponds to the strength of the prior. The relative values of the $\bm a$ vector corresponds to prior information on the comparative sizes of the parameters. 

\paragraph*{S3 Appendix.}
\label{S3_DAG}
{\bf Directed acyclic graph of the model} 

\begin{figure}[H]
\centering
\includegraphics{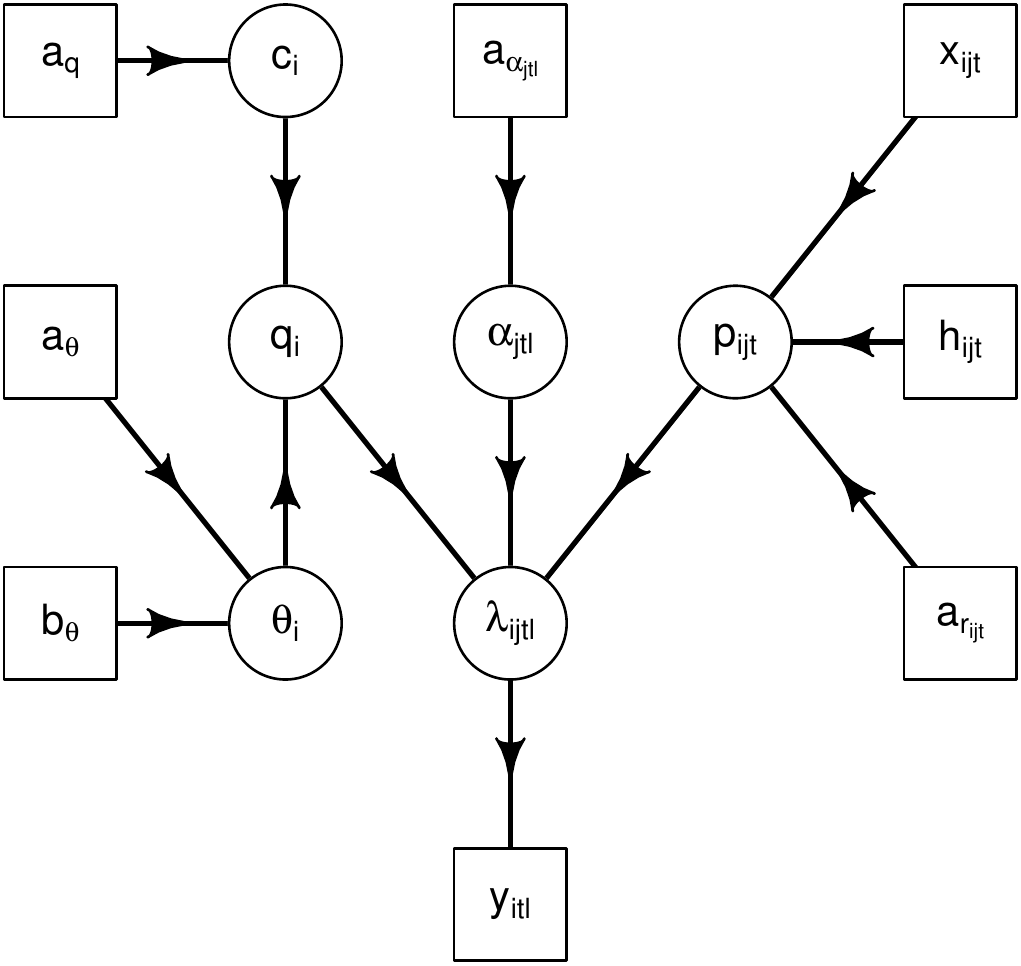}
\caption{Directed acyclic graph of the source attribution model. See Table \ref{table_params} for a concise description of the parameters.}
\label{fig:DAG}
\end{figure}

\paragraph*{S4 Appendix.}
\label{S4_type_source_effects}
{\bf Further details about the interpretation of the source and type effects.} 

The interpretation of source and type effects depends on the quality and type of data collected, the model specification, and the characteristics of the organism of interest. Source effects account for factors such as the amount of the food source consumed, the physical properties of the source and the environment provided for the bacteria through storage and preparation. 
Including an environmental source in the model can be thought of as grouping the (individually) unmeasured wildlife sources into one. It may also be a transmission pathway for pathogens present in livestock sources (for example, through the contamination of waterways) which complicates the interpretation meaning the source effects no longer directly summarise the ability of the source to act as a vehicle for food-borne infections \cite{HaldVosWed04}. Future work could involve attributing the water/ environmental samples to the other sources of infection (such as contamination from bovine, ovine, poultry, or other animal sources). Therefore, it would be possible to estimate the proportion of cases attributed to a sample directly, and via the environement.

\paragraph*{S5 Appendix.}
\label{S5_running_package}
{\bf Helpful details regarding use of \texttt{sourceR}} 

The \texttt{sourceR} package currently allows the relative prevalence matrix to be fixed at the maximum likelihood estimates, which includes zero values where a particular type was not detected in any samples from a source.  Fixing the relative prevalence matrix increases the posterior precision (and significantly reduces run time), but the results may be biased if the source data is not of high quality. Reducing the number of elements in the relative prevalence matrix $r$ that get updated at each iteration can significantly reduce computation time, however, the chains will converge more slowly. 

Care must be taken in performing marginal interpretations of the number of type parameters. It is much easier to split a group into two (with similar group means) than it is to merge two 
groups with clearly different means. Hence, a histogram of the number of groups per iteration is positively skewed compared to the true number of groups. When fitting the model with simulated data, visually assessing the dendrogram and heatmap to determine the number of groups usually provides a closer value to the true number of groups than looking at a histogram, particularly when the group means are well separated.

\paragraph*{S6 Appendix.}
\label{S6_mcmc_alg}
{\bf Full McMC Algorithm.} 
This section gives the full details of the algorithm used to fit our fully joint non-parametric source attribution model.  The outline McMC is shown in Algorithm \ref{alg:mcmc}. The Dirichlet distributed source effects $\bm{\alpha}_{tl}$ across times $t$ and locations $l$ (Step 1), and the relative prevalences $\bm{r}_{jt}$ across sources $j$ and times $t$ (Step 2) are updated using a constrained adaptive multisite logarithmic Metropolis-Hastings update step for 95\% of proposals, and a constrained adaptive multisite Metropolis-Hastings update step for the remainder to prevent the chain getting stuck at very low values \cite{RobRos06}. The adaptive algorithm updates the tuning value every 50 updates of the parameter.  This is further explained in Algorithm \ref{alg:constrainedMRW}.    

\begin{algorithm}
  \KwData {Human cases $\bm{y}$, source isolates $\bm{X}$, source prevalence $\bm{s}$}
  \BlankLine
  Initialize all parameters \;
  \For{$z$ times}{
    \ForEach{$t$, $l$}{
      \lnl{Step1} Update $\bm{\alpha}_{tl}$ \;
    }
    \ForEach{$j$,$t$}{
      \lnl{Step2} Update $\bm{r}_{jt}$ \; 
    }
    \lnl{Step3} Update $\bm{q}$ \; 
    Save chain state \;
  }
  \caption{\label{alg:mcmc}Outline McMC algorithm for the \texttt{HaldDP} model.}
\end{algorithm}

\begin{algorithm}
  \KwIn {$d$-dimensional Dirichlet$(\bm{a})$ distributed random variable $\bm{W}$, tuning variance vector $\bm{\sigma}$, online acceptance rate vector $\bm{\rho}$, $z$ the current McMC iteration number.}
  \KwOut {Updated $\bm{W}$ and $\bm{\sigma}$.}
  \BlankLine
  Let $\bm{W}^\prime = \bm{W}$ \;
  \For{$h$ times}{
    \lnl{} Let $j \sim \mbox{UniformInteger}[1, d]$ \;
    \lnl{} Let $g \sim \mbox{Uniform}[0, 1]$\;
    \If{g $> 0.05$}{
    Simulate $W^\prime_j = W_j * \exp \left[ \mathrm{N}(0, \sigma_j) \right]$
    
    $\delta = \frac{\bm{W}^\prime}{\bm{W}}$ 
    }
    \Else{Simulate $W^\prime_j = \mathrm{N}(W_j, 0.1)$
    
    $\delta = 1$}
    \lnl{Step 1} Let $\bm{W}^\prime = \bm{W}^\prime / | \bm{W}^\prime |$ \;
    \lnl{} Accept $\bm{W} = \bm{W^\prime}$ with probability
    $ 1 \wedge \frac{ f(\bm{W}^\prime | \bm{a}) }{ f(\bm{W} | a) } \cdot \delta $ and update $\rho_j$ \;
    \BlankLine
    \lnl{Step 2} \If{$h \mod 50 = 0$}{
      \If{$\rho_j > 0.44$}{$\sigma_j = \exp\left[\log (\sigma_j) + \left( 0.05 \wedge \frac{1}{\sqrt(z)} \right) \right]$}
      \Else{$\sigma_j = \exp\left[\log (\sigma_j) - \left( 0.05 \wedge \frac{1}{\sqrt(z)} \right) \right]$}
      }
  }
  \caption{\label{alg:constrainedMRW}Constrained adaptive multisite logarithmic random walk used for Dirichlet-distributed random variables.}
\end{algorithm}

For the Dirichlet process prior on $\bm{q}$, a marginal Gibbs sampler is constructed, as described in Algorithm \ref{alg:DPmarginalGibbs}.  Let $\mathcal{H}$ denote a set of cluster identifiers, with the $n$-dimensional group assignment vector $\bm{c}$ associating elements of $\bm{q}$ with clusters, such that $c_i = h$ assigns $q_i$ to cluster $h$.  Furthermore, each cluster $h$ assumes a value $\theta_h$ such that $q_i = \theta_{c_i}$.   

In Step 1 of Algorithm \ref{alg:DPmarginalGibbs}, conjugacy between the Gamma-distributed base distribution $P_0$ and the Poisson data likelihood permits the calculation of Multinomial conditional posteriors for elements of $\bm{c}$ arising from the Chinese Restaurant Process construction.  Here, the conditional posterior probability of type $i$ being assigned to group $h$ is as shown in Algorithm \ref{alg:DPmarginalGibbs}, with conjugacy permitting marginalisation with respect to the base distribution in order to calculate the probability of being assigned to a new group $h^\star$
\begin{equation*}
p_{h^\star}  = a_q \int_{\Theta} L(y_i | \theta, \lambda_i^\star)dP_0(\theta) = \frac{b_\theta^{a_\theta}(a_\theta + y_i)}{\Gamma (a_\theta) (b_\theta + \lambda^\star_i)^{a_\theta + y_i}}
\end{equation*}
with $y^\star_i = \sum_{t,l} y_{itl}$ and $\lambda^\star_i = \sum_{t,l} \bm{\alpha}_{tl}^T (\bm{r}_{it} \odot k_t)$

If a type is assigned to a new group, the set $\mathcal{H}$ is augmented and a corresponding cluster value is drawn from the posterior of $\theta_{h^\star}$.  Conversely, $\mathcal{H}$ is shrunk if a particular group becomes empty.  

In Step 2, the group values are drawn from the  posterior, conditional on $\bm{c}$.  The algorithm therefore alternates between updating group assignments $\bm{c}$ and group values $\bm{\theta}$. Hence, it explores the number of groups present, the type effects assigned to each group, and the values of each group.

\begin{algorithm}
\caption{\label{alg:DPmarginalGibbs}Marginal Gibbs sampling algorithm using the Chinese Restaurant Process construction of a Dirichlet process}

\KwData {Human case counts $\bm{y^\star} = \sum_{t,l} \{y_{1tl},\dots,y_{ntl}\}$, source~intensities~$\bm{\lambda^\star} : \lambda_{i}^\star = \sum_{t,l} \bm{\alpha}_{tl}^T (\bm{r}_{it} \odot k_t)$}
\KwIn {$\mathcal{H}$ the set of cluster identifiers,
 $\bm{c}$ an $n$-dimensional vector of group allocators, $c_i \in \mathcal{H}$,
 $\bm{\theta}$ a $|\mathcal{H}|$-dimensional vector of cluster values}
\BlankLine
\tcp{Update group allocation $c$}
 \For{$i$ in $1:n$}{
   \lnl{} Sample $c_i$ from $k(c_i | \cdot) \sim \mbox{Multinomial}(\left< p_h : h \in \mathcal{H}, p_{h^\star} \right>)$ where 
   \begin{align} p_h & = |\mathcal{H}_h^{(-i)}| L(y^\star_i | \theta_h, \lambda_i^\star),& & h \in \mathcal{H} \label{eq:pgroup} \\
    p_{h^\star} & = a_q \int_{\Theta} L(y^\star_i | \theta, \lambda_i^\star)dP_0(\theta),& & h \not\in \mathcal{H}
    \end{align}
     \;
   \If{$c_i = h^\star$}{
     Set $\mathcal{H} = \{\mathcal{H},h^\star$\} \;
     Sample $\theta_{h^\star} \sim  \mbox{Gamma}(y^\star_i + a_\theta, 1 + b_\theta)$ \;
   }
   \ElseIf{$|\mathcal{H}_h| = 0$}{
     Set $\mathcal{H} = \mathcal{H}^{(-h)}$ \;
   }
 }
  \tcp{Update cluster values $\bm{\theta}$}
  \For{$h$ in $\mathcal{H}$}{
  	\lnl{}  Update $\theta_h \sim \mbox{Gamma}(\sum_{i:c_i=h} y^\star_i + a_\theta, n_h + b_\theta)$
  }
  
\end{algorithm}

\paragraph*{S7 Appendix.}
\label{S7_cor_plot}
{\bf Posterior correlations and non-identifiability of source attribution.} 

\paragraph*{S8 Appendix.}
\label{S8_sim_study}
{\bf Worked example showing features of package using simulated data.} 

In this section, we provide a worked example using simulated data with multiple times and locations for source attribution data generated from the model in Section \ref{sec:model} (available in the \texttt{sourceR} data sets). There are two times (1, 2) and two locations (A, B) over which the human cases vary. The data must be in long format, with columns giving the number of human cases for each type, a column for each of the sources giving the number of positive samples for each type, and columns giving the time, location and type id's for each observation.  Note, the source data is the same for all locations within a time.

\begin{figure}[H]
\centering{
\includegraphics[width=0.8\textwidth]{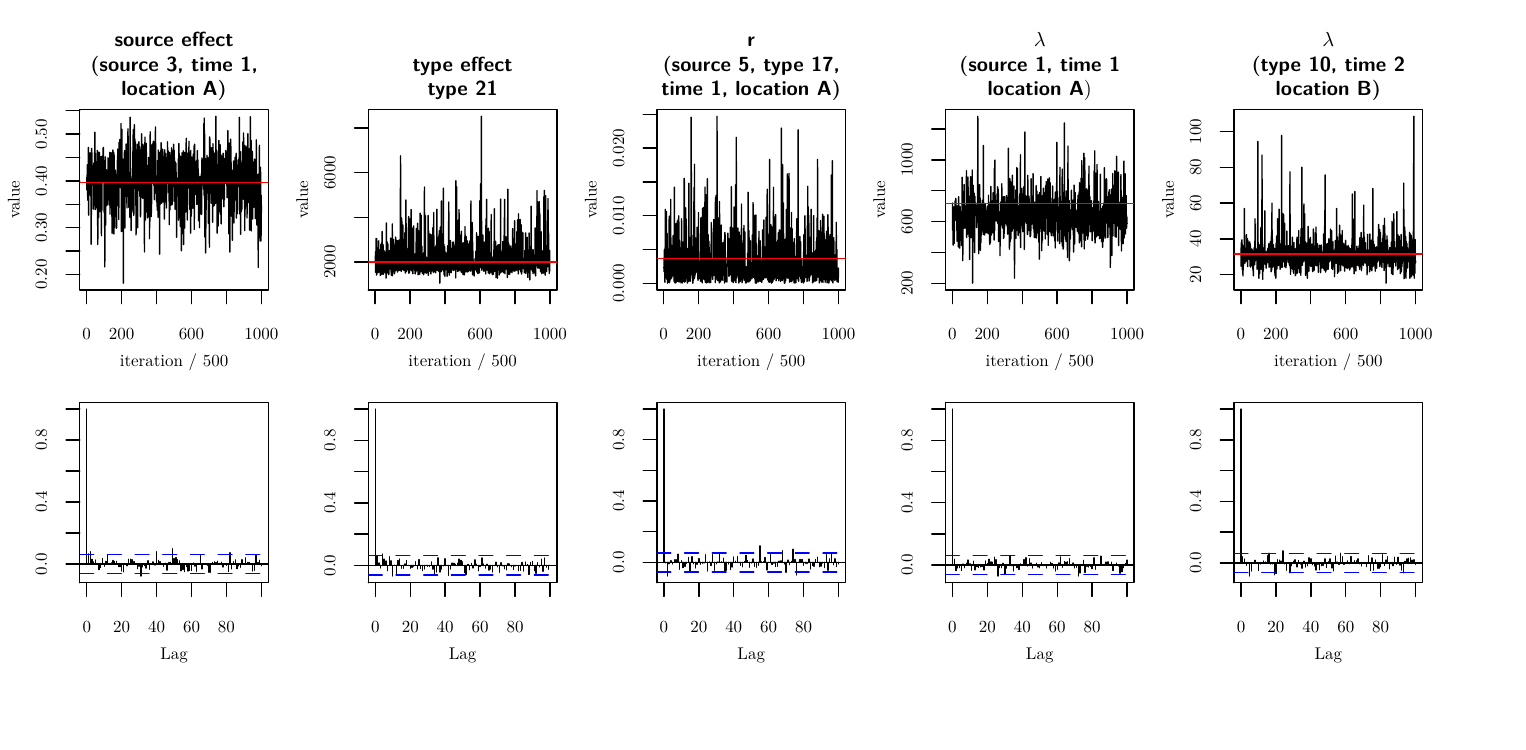} 
\caption{Trace and acf plots for a sample of the model parameters. True values of the parameters are shown in red.}\label{fig:trace_acf_sim_data_plots}}
\end{figure}

The algorithm is run for a total of 500,000 iterations (with a burn in of 10000 iterations and thinning 500). The acceptance rates for all parameters (except those updated using a Gibbs sampler) can be accessed using the \texttt{my\_model\$acceptance()}.
\begin{lstlisting}[frame=single]
## source and human case data
data(sim_SA_data)
## prevalences for each source/ time
data(sim_SA_prev)
## true values for the model parameters
data(sim_SA_true)
priors <- list(a_theta = 0.01, b_theta = 0.00001,
	           a_alpha = 1, a_r = 0.1)
my_model <- HaldDP$new(data = sim_SA_data, k = sim_SA_prev, 
                       priors = priors, a_q = 0.1)
\end{lstlisting}
Fitting parameters for the McMC are be passed using 
\begin{lstlisting}[frame=single]
my_model$fit_params(n_iter = 1000, burn_in = 2000, thin = 500)
\end{lstlisting}
The model is then run using
\begin{lstlisting}[frame=single]
set.seed(59623)
my_model$update()
\end{lstlisting}
Trace and autocorrelation plots for the parameters (Figure~\ref{fig:trace_acf_sim_data_plots}) indicate that the Markov chain is mixing well and has converged, and that thinning by 500 is adequate. The following R code demonstrates how to access and plot the marginal posteriors for some parameters.
\begin{lstlisting}[frame=single]
## Plot the marginal posterior for source effect 2, time 1, location A
plot(my_model$extract(params = "alpha", times = "1", locations = "B", 
     sources = "Source4")$alpha, type="l")
     
## Plot the marginal posterior for the type effect 21
plot(my_model$extract(params = "q", types = "21")$q, type="l")

## Plot the marginal posterior for the relative prevalence of 
## source effect 5, type 17, at time 2
plot(my_model$extract(params = "r", times = "2", sources = "Source5", 
     types = "17")$r, type="l")
     
## Plot the marginal posterior for lambda_j source 1, time 1, location A
plot(my_model$extract(params = "lambda_j", times = "1", locations = "A", 
     sources = "Source1")$lambda_j, type = "l")
     
## Plot the marginal posterior for lambda_i 10, time 2, location B
plot(my_model$extract(params = "lambda_i", times = "2", locations = "B", 
     types = "10")$lambda_i, type="l")
\end{lstlisting}

\begin{figure}[H]
\centering{
\includegraphics[width=0.8\textwidth]{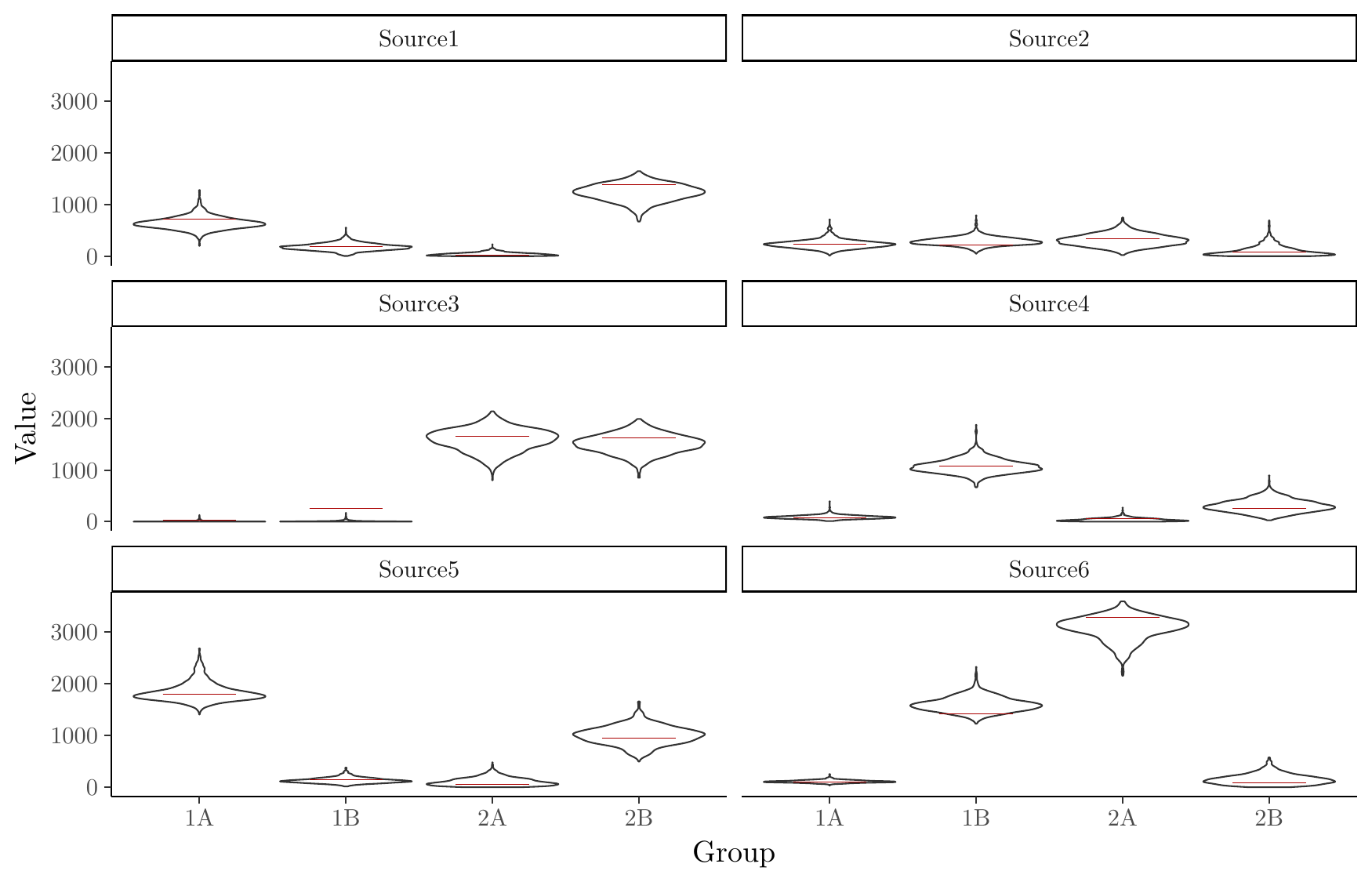} 
\caption{Violin plots showing marginal posteriors for each $\lambda_j$ (number of cases attributable to each source) for each time (1, 2) and location (A, B). True $\lambda_j$ values are shown as horizontal red lines.}\label{fig:sim_pois_lambdaj_plots}}
\end{figure}

\begin{figure}[H]
\centering{
\includegraphics[width=0.8\textwidth]{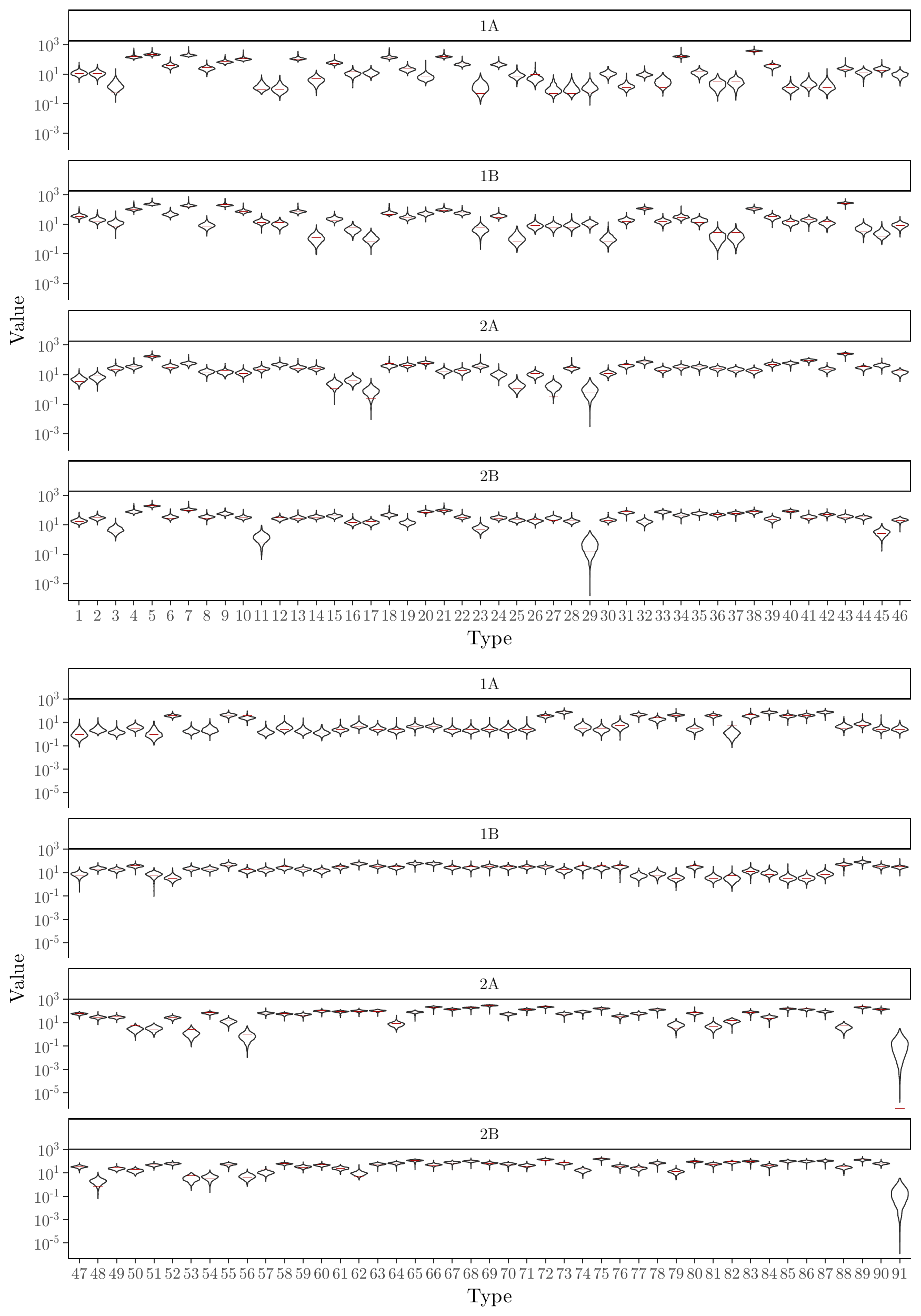} 
\caption{Violin plots showing the marginal posteriors for each $\lambda_i$ (number of cases attributed to each type) for each time (1, 2) and location(A, B).
True $\lambda_i$ values are shown as horizontal red lines.}\label{fig:lambda_i_residuals}}
\end{figure}

\begin{figure}[h]
\centering{
\includegraphics[width=0.8\textwidth]{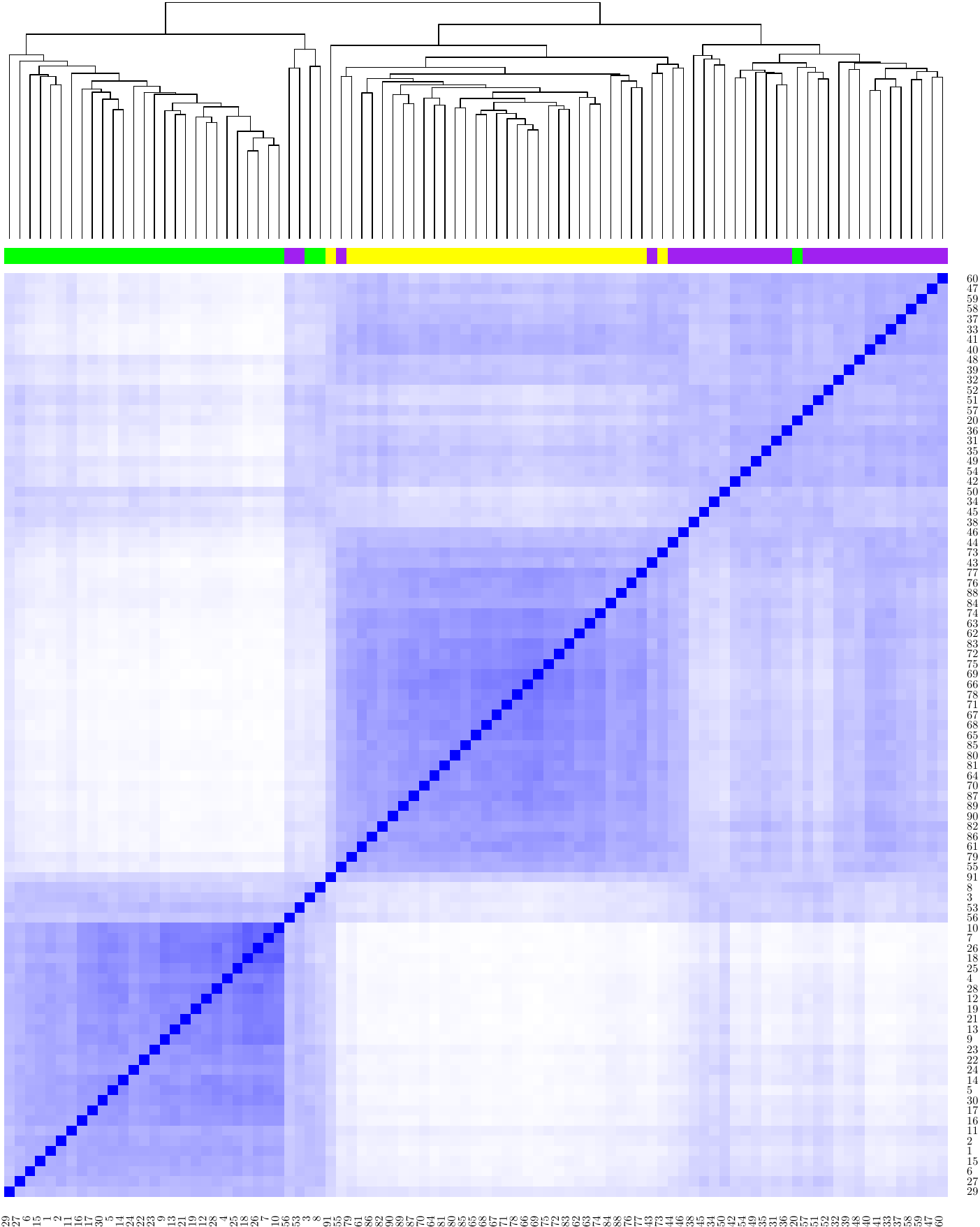} 
\caption{Heatmap showing the grouping of the type effects (q) using simulated data (true groupings given by the 3 colours in the bar under the 
dendrogram).}\label{fig:type_effect_heatmap}
}
\end{figure}

Medians and credible intervals can be obtained for each parameter using \texttt{res\$summary()}. 
The marginal density plots of the number of cases attributed to each source at each time and location ($\lambda_{jtl}$) show that the true values (shown by a red horizontal line on the graph) are being estimated well (Figure~\ref{fig:sim_pois_lambdaj_plots}). The violin plots of the number of cases attributed to each type (residual plot) for $\lambda_{i}$ (Figure~
\ref{fig:lambda_i_residuals}) shows that the model is fitting well. The heatmap shows the grouping of the type effects (Figure~\ref{fig:type_effect_heatmap}) computed using a dissimilarity matrix from the clustering output of the McMC. The coloured bar under the dendrogram gives the correct grouping from the simulated data. This shows that 
the majority of types have been classified correctly.

\section{Acknowledgments}
The research for this paper was financially supported by the Ministry for Primary Industries, the Institute of Fundamental Sciences (Massey University), the mEpiLab (Massey University), and CHICAS (Lancaster University). We acknowledge the following individuals and groups: mEpiLab
(Massey University), MidCentral Public Health Services and Petra Mullner (for the Manawatu data set) and Geoff Jones (for his helpful input on automatic clustering methods).


\end{document}